\documentclass[structabstract]{aa}
\usepackage{graphicx}
\usepackage{txfonts}

\newcommand{\msun}{M_\odot}
\newcommand{\mbh}{M_\bullet}
\newcommand{\mcnd}{M_\mathrm{CND}}
\newcommand{\rcnd}{R_\mathrm{CND}}

\newcommand{\pc}{\mathrm{pc}}
\newcommand{\rd}{\mathrm{d}}
\newcommand{\bn}{\mathbf{n}}
\newcommand{\bs}{\!\!\!}
\newcommand{\NEW}[1]{{#1}}

\begin{document}

\title{The warped young stellar disc in the Galactic Centre}

\author{L. \v{S}ubr\inst{1,2,3}\thanks{E-mail: subr@sirrah.troja.mff.cuni.cz},
J. Schovancov\'a\inst{1} \and P. Kroupa\inst{3}}

\institute{$^1$Faculty of Mathematics and Physics, Charles University,
V Hole\v{s}ovi\v{c}k\'ach 2, CZ-18000 Praha, Czech Republic\\
$^2$Astronomical Institute, Academy of Sciences, Bo\v{c}n\'{\i}~II, CZ-14131~Praha,
Czech Republic\\
$^3$Argelander Institute for Astronomy (AIfA), Auf dem H\"ugel 71, D-53121 Bonn,
Germany}
\date{Accepted .... Received ....}



\abstract{}
{Within the central parsec of the Galaxy, several tens of young stars orbiting
a central supermassive black hole are observed. A subset of these
stars forms a coherently rotating disc. Other observations reveal
a massive molecular torus which lies at a radius $\sim1.5\mathrm{pc}$ from the
centre. In this paper we consider the gravitational influence of the molecular
torus upon the stars of the stellar disc.}
{We derive an analytical formula for the rate of precession of individual
stellar orbits and we show that it is highly sensitive upon the orbital
semi-major axis and inclination with respect to the plane of the torus as well
as on the mass of the torus.}
{Assuming that both the stellar disc and the molecular torus are stable on the
time-scale $\gtrsim6 \mathrm{Myr}$, we constrain the mass of the
torus and its inclination with respect to the young stellar disc.
We further suggest that all young stars observed in the Galactic Centre may
have a common origin in a single coherently rotating structure with an opening
angle $\lesssim 5\degr$, which was partially destroyed (warped) during its
lifetime by the gravitational influence of the molecular torus.}{}


\maketitle

\section{Introduction}
Near infrared observations of the central parsec of the Galaxy that were
made over the past decade have brought new views of the environment
in the vicinity of a supermassive black hole. They revealed a numerous
population of young massive stars which may be distinguished into at least
two different groups: within a distance $\lesssim0.03\mathrm{pc}$ from
the centre there are found more than ten
so called S-stars orbiting the supermassive black hole on apparently
randomly oriented and highly eccentric ($e\gtrsim0.8$) orbits. They appear to
be standard OB main sequence stars (Ghez et al.~2003, Eisenhauer et al.~2005)
which is in contradiction with strong tidal forces that prevent stellar
formation at the place. Unless these stars mimic their age, it is likely
that they have migrated to the centre from larger distances.

Further away, at $0.03\mathrm{pc} \lesssim r \lesssim 0.5\mathrm{pc}$, nearly
up to one hundred young stars were detected (see Paumard et al.~2006 for
one of the most recent reviews). These stars are classified mainly as post
main sequence OB supergiants and Wolf-Rayet stars. According to the
evolutionary phase, their age is estimated to be $6\pm2\mathrm{Myr}$.
Levin \&~Beloborodov~(2003) pointed out that
a substantial fraction of these stars form a coherently rotating disc (usually
referred to as a `clockwise' stellar disc or CWS).
It is a flaring disc with an opening angle $\approx15\degr$ with a rather
sharp inner edge at $0.03\pc$ and it extends up to radius of $\approx0.3\pc$.
The radial column density profile of the CWS decreases approximately
as $r^{-2}$, i.e. most of the stars are concentrated at the inner edge.
The mean plane of the disc can be determined by two angles: inclination
$i^\prime\approx127\degr$ with respect to the plane of the sky and longitude
of the ascending node $\Omega^\prime\approx99\degr$ (measured from the
North direction; see Paumard et al.~2006 for a detailed description of the
convention). Levin \& Beloborodov~(2003) suggested that this disc-like pattern
is a consequence of a stellar formation in a self-gravitating accretion disc.

Further analyses (Genzel et al.~2003, Paumard et al.~2006) indicate the
presence of another coherent stellar system which is usually referred to as
the `counter-clockwise' stellar disc (CCWS). This structure is
assumed to be formed by fewer ($\lesssim15$) stars, it is narrower in the
radial extent being concentrated around $r\gtrsim0.15\pc$ and has a larger
opening angle $\approx20\degr$. The existence of the CCWS disc is a matter
of an ongoing debate (e.g. Lu et~al.~2007), nevertheless, even if it is
accepted as an explanation
of the origin of another subset of young stars in the Galactic Centre, there
would still remain more than twenty stars not belonging to any of the two
stellar discs and, therefore, without a satisfactory explanation of their
origin.

The gravitational potential in the considered region is dominated by the
supermassive black hole of mass $\mbh\approx3.5\times10^6\msun$ (Ghez et
al.~2003). It is surrounded by a roughly spherical cluster of late-type stars.
The radial density profile is well fitted with a broken power-law
$\rho(r)\propto r^{-\beta}$ with index
$\beta=1.19$ below $r=0.22\pc$ and $\beta=1.75$ above the break radius
(Sch\"odel et al.~2007).
Its mass, $M_\mathrm{c}$ within $1\pc$ is comparable to the mass of the
black hole.

The central region is surrounded by a molecular torus
(circum-nuclear disc; CND)
which lies at the outer edge of the black hole's sphere of influence
($\rcnd\approx 1.5\pc$). Its mass estimated from the radio observations of
ionised molecular gas is $\mcnd\approx10^6\msun$ (Christopher et al.~2005).
This massive structure defines a non-spherical component of the gravitational
field in the central parsec.

In this paper we investigate the influence of the CND
upon the dynamical evolution of the disc-like stellar structures. In the
subsequent section we briefly review the dynamics in the
perturbed Keplerian potential. In Sec.~\ref{sec:results} we apply the
results on the motion of the stellar discs --- we present constraints on
some parameters of the CND and the CWS determined from their gravitational
interaction and we also give suggestions on the dynamics of the whole
system of young stars over its lifetime. Conclusions and discussion of
our results are given in Sec.~\ref{sec:conclusions}.

\section{Dynamics in the perturbed Keplerian potential}
For the purpose of the study presented in this paper we introduce a simple
model of the Galactic Centre which consists of three main constituents
determining the gravitational field: (i) the central supermassive black hole
of mass $\mbh=3.5\times10^6\msun$ which is treated as a source of the Keplerian
potential, (ii) the massive molecular torus modelled as an infinitesimally thin
ring of radius $\rcnd$ and mass $\mcnd$ and (iii) a spherical stellar cusp with
a power-law density profile $\beta$ and mass $M_\mathrm{c}$ within the radius
$\rcnd$. Both the ring and the cusp are centred on the black hole.

The stars are treated as test particles whose motion is determined by
the composed smooth external potential. Their orbits can be represented
by five orbital elements: the semi-major axis, $a$, eccentricity, $e$,
inclination, $i$, argument of pericentre, $\omega$, and longitude of the
ascending node, $\Omega$. Here we assume the angles to be measured
in the frame in which the ring lies in the $x$-$y$ plane. For convenience,
the results presented in the subsequent section will be transformed into
coordinates with $x^\prime$-$y^\prime$ representing the plane of the sky
and $z^\prime$ pointing from the observer. The inclination and longitude of
the ascending node in this observer's coordinate
system will be denoted $i^\prime$ and $\Omega^\prime$, respectively.

If the gravity of the spherical cusp were ignored, the dynamics in the
field of the central body and the ring would be equivalent to the reduced
hierarchical three body problem. In this case, the orbital elements $e,\;i,
\;\omega$ and $\Omega$ will undergo secular evolution (Kozai~1962,
Lidov~1962) on a time-scale of
\begin{equation}
T_\mathrm{K} \equiv\frac{\mbh}{\mcnd} \, \frac{\rcnd^3}{a\sqrt{G\mbh a}}\;.
\label{eq:TK}
\end{equation}
The equations of motion for {\em mean\/} orbital elements read:
\begin{eqnarray}
T_\mathrm{K}\,\sqrt{1-e^2}\,\,\frac{\rd e}{\rd t} &\bs=\bs&
 {\frac{15}{8}}\,e\,(1-e^2)\,\sin2\omega\,\sin^{2}i\,,
 \label{eq:dedt} \\
T_\mathrm{K}\,\sqrt{1-e^2}\,\,\frac{\rd i}{\rd t} &\bs=\bs&
 -\frac{15}{8}\,e^2\,\sin2\omega\,\sin i\,\cos i\,,
 \label{eq:didt} \\
T_\mathrm{K}\,\sqrt{1-e^2}\,\,\frac{\rd\omega}{\rd t} &\bs=\bs&
 \frac{3}{4}\left\{ 2-2e^2+5\sin^{2}\omega\left[e^{2}-\sin^{2}i\right]\right\}\,,
\label{eq:dodt} \\
T_\mathrm{K}\,\sqrt{1-e^2}\,\,\frac{\rd\Omega}{\rd t} &\bs=\bs&
 -\frac{3}{4}\cos i \left[1+4e^2-5e^2\cos^{2}\omega\right]\,.
\label{eq:dOdt}
\end{eqnarray}
The temporal evolution does not depend upon the angle $\Omega$ which is
merely a consequence of the axial symmetry; furthermore, energy conservation
implies a constant $a$ in this order of approximation.

\begin{figure}
\includegraphics[width=\columnwidth]{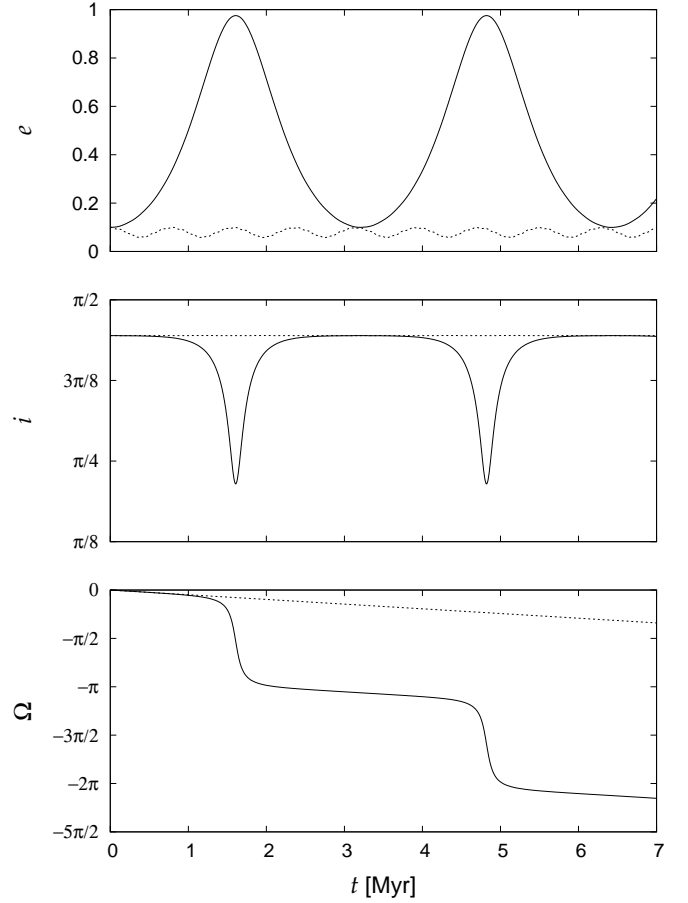}
\caption{Evolution of orbital elements of two example orbits. The solid line
represents a trajectory in the gravitational field of the central mass
$\mbh=3.5\times10^6\msun$ and a ring of radius $\rcnd=1.5\mathrm{pc}$
and mass $\mcnd=\mbh$. The dotted line shows an orbit
integrated in the field including in addition a spherical cusp of mass
$M_\mathrm{c}=0.1\mbh$. In both cases the initial values of the orbital
elements are: $a=0.1\rcnd,\, e=0.1,\, i=80\degr$ and $\Omega=0$.}
\label{fig:trajectory}
\end{figure}

Including the gravity of the spherical cusp leads to an additional shift of
the pericentre which can be incorporated by an extra term in
equation~(\ref{eq:dodt}),
dependent upon the global parameters of the cusp and the semi-major axis
and eccentricity of the orbit (Ivanov et al.~2005). The overall influence
of the cusp can be characterised by a decrease of the amplitude of the
oscillations of eccentricity and inclination and shortening of their period
(Karas \& \v{S}ubr~2007). This is clearly seen also in
Fig.~\ref{fig:trajectory} which shows the evolution of an example orbit both
in the case with and without the potential of the spherical cusp.
On the other hand, a generic influence of the cusp upon the evolution of
$\Omega$ lies in diminishing the variations of its first time derivative;
the characteristic time-scale of the change of $\Omega$ becomes in general
much longer than that of the mutually coupled elements $e,\;i$ and $\omega$.

Within the context of this paper we are interested in a system where
$M_\mathrm{c} \gtrsim 0.1\mcnd$. In this case the amplitude of the oscillations
of eccentricity and inclination can be considered
negligible and $\omega$ rotates with frequency much higher than that of
$\Omega$. This configuration allows us to simplify eq.~(\ref{eq:dOdt})
by averaging over one revolution of $\omega$:
\begin{equation}
 \frac{\rd\Omega}{\rd t} \approx -\frac{3}{4}\, \frac{\cos i}{T_\mathrm{K}}\,
 \frac{1+\frac{3}{2}e^2}{\sqrt{1-e^2}} \approx const\,.
\label{eq:dOdt_approx}
\end{equation}
The change of $\Omega$ over an interval $\Delta t$ can then be written as:
\begin{eqnarray}
 \Delta \Omega &\bs=\bs& - \frac{3}{4}\, \cos i\; a^{3/2}\,
 \frac{\sqrt{G\mbh}}{\rcnd^3} \frac{\mcnd}{\mbh}\,
 \frac{1+\frac{3}{2}e^2}{\sqrt{1-e^2}}\, \Delta t \nonumber \\
 &\bs=\bs& -5417\degr\; \left( \frac{\mbh}{3.5\times10^6\msun} \right)^{1/2}
 \left( \frac{\rcnd}{1\mathrm{pc}} \right)^{-3/2}
 \label{eq:dO} \\
 & & \times\, \cos i\, \frac{1+\frac{3}{2}e^2}{\sqrt{1-e^2}}\,
 \left( \frac{a}{\rcnd} \right)^{3/2}
 \frac{\mcnd}{\mbh}\, \frac{\Delta t}{1\mathrm{Myr}}
 \nonumber
\end{eqnarray}
According to the underlying perturbation theory, eqs.~(\ref{eq:TK})
-- (\ref{eq:dO}) refer to the elements averaged over one orbital period,
which cannot be trivially mapped to the osculating elements defined
by instant positions and velocities. We have performed numerical tests
in order to check the error introduced by replacing averaged elements
by the osculating ones in formula~(\ref{eq:dOdt_approx}). For $M_\mathrm{c}
\sim \mbh$ and $e\lesssim0.5$ the difference between the analytical estimate
and the numerically measured rate of change of $\Omega$ was always smaller
than a factor of $1.2$. With increasing eccentricities,
formula~(\ref{eq:dOdt_approx}) overestimates the real precession rate
by a somewhat larger factor which, however, still stays smaller than $2$.

\NEW{The real source of the perturbing potential, CND, is rather clumpy torus of
a finite thickness. Hence, we have performed several numerical integrations
of a test particle orbit in a gravitational field of a central mass, spherical
cusp and a set of $\sim10$ discrete point masses with orbits confined in a
torus of a toroidal and poloidal radii $1.5\mathrm{pc}$ and $0.2\mathrm{pc}$,
respectively. The difference in the orbital evolution with respect to that in
the case of a ring-like perturbation was found to be negligible for orbits
with $a\lesssim0.5\mathrm{pc}$.}

\section{Consequences of differential precession}
\label{sec:results}
\subsection{Constraints on the CND}
\label{sec:results1}
Let us now consider an ensemble of stars forming a thin disc, i.e. with
inclinations and longitudes of ascending nodes lying in a narrow interval.
Let us further assume that the stellar orbits evolve solely due to the
external gravitational potential determined by fixed parameters $\mbh,\;
M_\mathrm{c}\approx\mbh,\;\mcnd\lesssim\mbh$ and $\rcnd \gg a$. The key
feature of the orbital evolution will be precession around the symmetry
axis of the ring-like component of the gravitational field.

If the semi-major axes of stars at the inner edge of the disc
are smaller than those at the outer edge by a factor of $\gtrsim5$, the
two edges of the disc will precess at a rate different by a factor
$\gtrsim10$. After a certain period of time, their angular momenta
will point to completely different directions, i.e. the disc-like structure
will be destroyed. Hence, the requirement of the stability of the disc
over a given period of time transforms into the requirement of a sufficiently
slow precession at its outer edge.

Let us consider the subset of the young stars in the Galactic Centre which
form the `clockwise' stellar disc (CWS).
Inserting values $\rcnd=1.5\mathrm{pc},\; a=0.1\rcnd,\; \Delta t=6\mathrm{Myr}$
and $e=0$ into eq.~(\ref{eq:dO}), we obtain
\begin{equation}
 \Delta\Omega = -560\degr\, \cos i\, \frac{\mcnd}{\mbh}\;.
\label{eq:dO_out}
\end{equation}
In order to be compatible with observations, $\Delta\Omega$ has to be smaller
than $\sim10\degr$, which is the opening angle of the inner part of the CWS
(Beloborodov et al.~2006).
Hence, eq.~(\ref{eq:dO_out}) poses constraint upon the inclination of the
disc with respect to the molecular torus, depending on its mass.
Considering e.g. $\mcnd\approx0.3\mbh$ (Christopher et al.~2005) requires
$\cos i < 0.06$, i.e. $i \in \langle 86\degr, 90\degr \rangle$.
Simultaneously, this poses an upper limit $\theta_0\lesssim5\degr$ on the
{\em initial\/} opening angle of the stellar disc. For the sake of simplicity,
we have considered a common sense of precession of all disc stars,
i.e. $i<90\degr$. Identical results would be obtained for $i>90\degr$ due
to the symmetry of the problem.
\NEW{The constraint would be tighter by a factor of $\approx 1.5$ if we
consider nonzero ($\sim0.5$) eccentricities of the stellar orbits.}

To conclude this analysis, we remark that
$\cos i = \mathbf{n}_\mathrm{CWS} \centerdot \mathbf{n}_\mathrm{CND}
< 0.06$ is in accord with estimates of the normal vectors\footnote{We
follow the convention of Paumard et al.~(2006) according to which angles
$i^\prime$ and $\Omega^\prime$ are related to the normal vector of the orbital
plane as: $(n^\prime_x, n^\prime_y, n^\prime_z) = (\sin i^\prime
\cos\Omega^\prime, -\sin i^\prime \sin\Omega^\prime, -\cos i^\prime)$.}
of the plane of the disc and torus
($\mathbf{n}^\prime_\mathrm{CWS}=(-0.12,\, -0.79,\, 0.60)$, Paumard et al.~2006;
$\mathbf{n}^\prime_\mathrm{CND}=(0.85,\, -0.40,\, -0.34)$,
Jackson et al.~1993).

\subsection{A common origin of young stars in the Galactic Centre?}
Formula~(\ref{eq:dO_out}) indicates that orbits of stars at radii
$\gtrsim0.1\mathrm{pc}$ and/or inclinations $i<85\degr$ or $i>95\degr$
were considerably affected by precession within the past $6\mathrm{Myr}$,
i.e their current orbital parameters are different from their values
at the time of the birth. We suggest a possibility that stars which are not considered to be
members of the CWS nowadays have been its members at the
time of its formation. During the $\sim6\mathrm{Myr}$ of the dynamical
evolution their orbits were subject to precession due to the gravity of the
CND and were detached from their parent stellar system. This model could
represent a possible solution of the problem of the origin of all young stars
in the Galactic Centre.

The mapping between the initial and the current orientation of the stellar
orbit is formally rather straightforward within our simple model. Unfortunately,
the observational data do not provide us with sufficiently accurate values of
the parameters $\rcnd$ and
$\mcnd$. Furthermore, the high sensitivity of the precession rate upon the
inclination and semi-major axis together with a lack of robust determination
of these orbital elements from the observational data also stand as a severe
obstacle in an attempt to track the orbits of the observed stars back in time,
which could prove or discard the hypothesis of a common origin. In the rest
of this section we describe a test shows that our model is compatible with the
publicly available observational data.

We have taken data from Table~2 of
Paumard et al.~(2006) from which we have considered all stars with determined
3D~velocity and index $\geq15$ (i.e. excluding the S-stars) which gives
in total $N_\star=72$ stars. Five free parameters of the model consist of the
two angles, $(\Omega^\prime_0, \, i^\prime_0)$, determining the initial
orientation, $\bn^\prime_0$, of the stellar disc; another two angles,
$(\Omega^\prime_\mathrm{CND},\, i^\prime_\mathrm{CND})$, determine
the orientation of the CND, and $\mcnd$ represents its mass. (The last
parameter can be considered as a degenerate combination
of $\mcnd,\;\rcnd$ and $\Delta t$; in the following, we will implicitly
assume $\rcnd=1.5\mathrm{pc}$ and $\Delta t = 6\mathrm{Myr}$.) For a given
set of parameters we scan the $1\sigma$ neighbourhood of each star's velocity
with sampling $d_v$. The $x^\prime$ and $y^\prime$ coordinates
of the stars' positions are assumed to be determined exactly. On the other
hand, the $z^\prime$ coordinate (along the line of sight) is unknown.
Therefore, we scan it with sampling $d_z$ in a whole range allowed by
the condition that the star is gravitationally bound to the black hole.
In total, we consider $V_{1\sigma}=d_z d_v^3$ pairs of position and velocity
vectors which represent states compatible with the observational data of a
particular star. For each state we perform a rotation of the normal vector of
the orbit around the axis of the CND according to formula~(\ref{eq:dO}), which
gives its direction, $\bn_{j,0}$ at $t=0$, i.e. $6\mathrm{Myr}$ ago. We
further calculate its angular distance to $\bn_{0}$, $\cos\delta_0 =
\bn_{j,0}\centerdot \bn_0$ and count the number of states, $N_{j,5}$, with
$\delta_0 < 5\degr$. We consider the measured star's position and velocity
to be compatible with the hypothesis that it was born in the disc with normal
vector $\bn^\prime_0$ and thickness $5\degr$, provided $N_{j,5}>0$. Finally,
we denote $N_5(\Omega^\prime_0, i^\prime_0,
\Omega^\prime_\mathrm{CND}, i^\prime_\mathrm{CND}, \mcnd)$ the number of stars
with $N_{j,5}>0$ for a given set of values of the parameters of the model.

\begin{figure*}
\includegraphics[width=\textwidth]{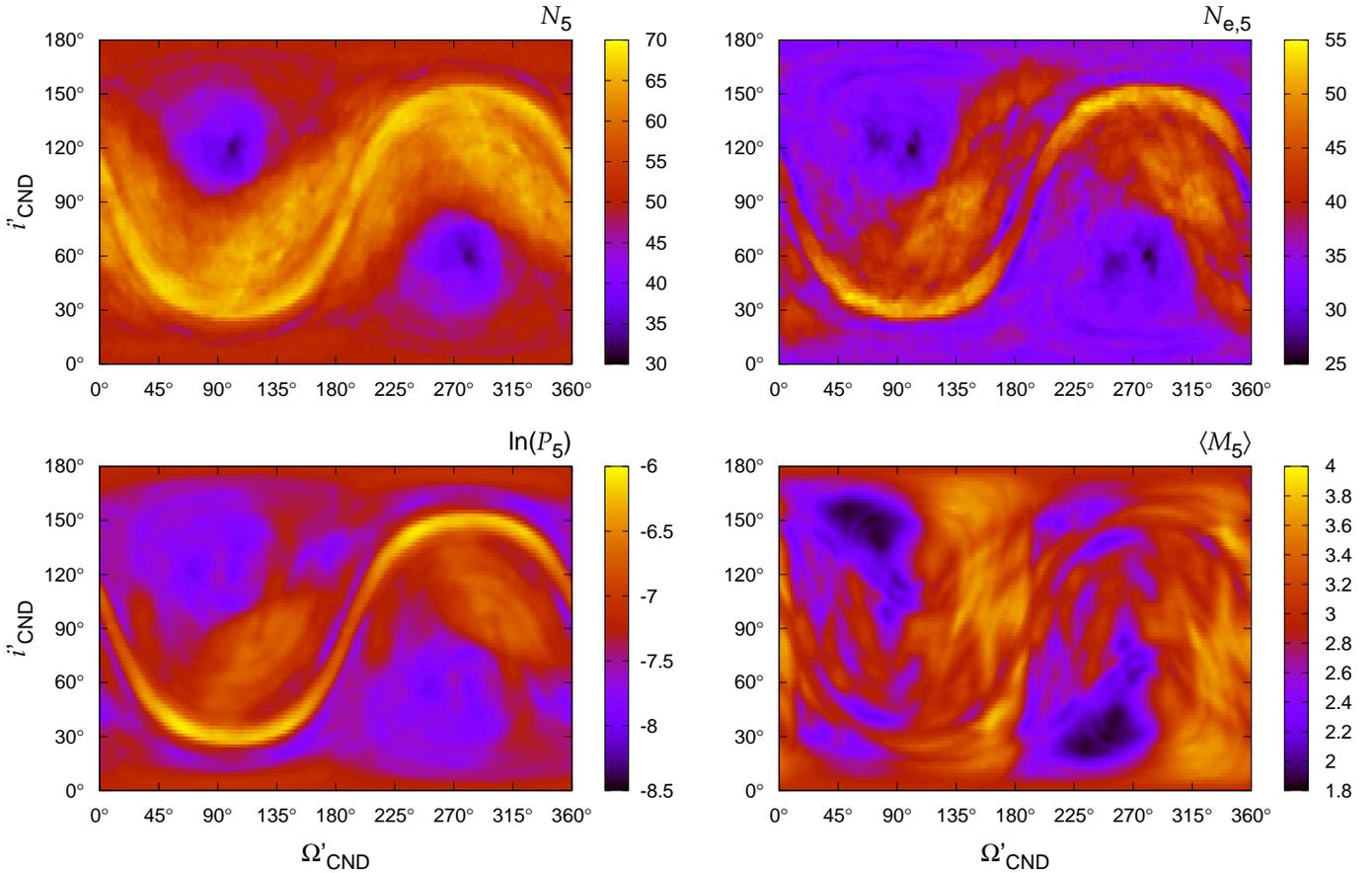}
\caption{Various tests of compatibility of the hypothesis of a common origin
of young stars in a single thin disc. $N_5$: number of stars that have at least
one state in the $1\sigma$ neighbourhood of observed velocities that corresponds
to an orbit dragged from the parent disc with normal vector
$(\Omega^\prime_0,\, i^\prime_0) = (99\degr,\, 120\degr)$ by the CND of mass
$\mcnd=\mbh$. $N_{e,5}$: similar to the previous but excluding orbits with
$e>0.5$. ${\cal P}_5$: volume of the subset of the $1\sigma$ neighbourhood
occupied by the states originating in the parent disc. \NEW{$\langle M_5 \rangle$:
mean value of the mean anomaly of orbits with $\delta_0 < 5\degr$.}}
\label{fig:test}
\end{figure*}

As we have discussed in Section~\ref{sec:results1}, the inner part of the
CWS must have undergone only negligible precession due to the gravity
of the CND. Therefore, we assume that it conserves orientation of the putative
single parent disc and we consider $(\Omega^\prime_0,\, i^\prime_0) =
(99\degr,\, 120\degr)$ which is the normal vector of the inner part
of the CWS according to Beloborodov et al.~(2006). We further set $\mcnd=\mbh$
which enables us to plot $N_5$ as a function of $\Omega^\prime_\mathrm{CND}$
and $i^\prime_\mathrm{CND}$ as it is shown in Figure~\ref{fig:test}. We
see that there exists an extended region where the observational data of
nearly all stars are compatible with the hypothesis of their origin in a
parent disc of thickness $\sim5\degr$.
This region of large values of $N_5$ extends along the set of
$(\Omega^\prime_\mathrm{CND},\, i^\prime_\mathrm{CND})$ perpendicular to the
normal vector $\bn^\prime_0$. This is a natural consequence
of the assumption that $\sim35$ stars, identified as CWS nowadays, haven't
undergone large precession. The region of good compatibility also includes
an approximate orientation of the CND as determined from observations,
$(\Omega^\prime_\mathrm{CND},\, i^\prime_\mathrm{CND})
= (25\degr, 70\degr)$, e.g. by Jackson et al.~(1993).

We have performed
analogical test of compatibility also for $\mcnd=0.3\mbh$ and $3\mbh$ and
$\sim10\degr$ neighbourhood of $(\Omega^\prime_0,\, i^\prime_0) =
(99\degr,\, 120\degr)$. In all cases we have obtained a picture similar in that
there exists an extended region with $N_5\gtrsim68$. Enlarging the inspected
neighbourhood of the observed velocities to $3\sigma$ leads to larger values
of $N_5$ with its maximum reaching $72$.
This means that the observational data
are compatible with the hypothesis of the common origin of the young stars in
a single thin disc, nevertheless, they
do not pose strong constraints on the parameters of the model.
We introduce three supplementary tests
that may be applied to the observational data to verify validity of our
hypothesis. First, the model of star formation is assumed to prefer low
eccentricities of the stellar orbits. Hence,
we introduce $N_{e,5}$ in the same way as $N_5$, but now with an additional
condition $e<0.5$ for all tested orbits. Middle panel of Fig.~\ref{fig:test}
shows that region with large $N_{e,5}$ coincides with the region of large $N_5$.
Maximum value of $N_{e,5}$ is $\sim55$, i.e the model requires about one third
of the stellar orbits to have moderate to large eccentricities. Again,
considering $3\sigma$ neighbourhood of the measured velocity vectors weakens
this constraint, giving $N_{e,5}\gtrsim70$ for a wide range of the model
parameters. Additional
analysis reveals that most of the eccentric orbits does not belong to the
CWS subset of stars.

A ratio $N_{j,5}/V_{1\sigma}$ can be
considered as a measure of the
probability that the orbit of star $j$ originated in a disc with opening
angle $5\degr$ and normal vector $\approx \bn_{0}$. Consequently,
we introduce
\begin{equation}
 {\cal P}_5 \equiv \left( \prod_{j=1}^{N_\star}
 \frac{N_{j,5} + 1}{V_{1\sigma}} \right)^{1/N_\star}
\label{eq:pst}
\end{equation}
as a measure of the probability that all stars originated in a thin disc
(Adding a unity to $N_{j,5}$ in~(\ref{eq:pst}) prevents ${\cal P}_5$ from
being zero everywhere while it does not strongly affect its meaning.)
Bottom panel of Fig.~\ref{fig:test} shows that ${\cal P}_5$ accentuates
orientations of the CND nearly perpendicular to the normal vector
$(\Omega^\prime_0,\, i^\prime_0)$, but it does not strongly discriminate
among the models that fall into this region. Maximum of this function is
at $(\Omega^\prime_\mathrm{CND},\, i^\prime_\mathrm{CND}) \approx (55\degr,\,
40\degr)$.

\NEW{Following Beloborodov et~al.~(2006), the configurations compatible with
the hypothesis of a single warped disc are expected to have equally distributed
value of the mean anomaly, $M$, of the individual orbits. Full test of the
distribution of the mean anomaly of all configurations that have fulfilled
other criteria of compatibility is not possible as it would require analysis
of $\approx {\cal P}_5 V_{1,\sigma}^{N_\star}$ {\em combinations\/} of orbits.
(Note, that other tests presented here require analysis of only $N_\star V_{1,\sigma}$ individual orbits.) In the bottom panel of Fig.~\ref{fig:test}
we present a restricted test showing the mean value of the mean anomaly,
$\langle M_5 \rangle$, for {\em all\/} tested orbits with $\delta_0 < 5\degr$.
This quantity is close to $\pi$ which corresponds to the expected uniform
distribution in the major part of the $(\Omega^\prime_\mathrm{CND},\,
i^\prime_\mathrm{CND})$ space.
Analogical plot would show that $\langle M^2_5 \rangle$ is close to
the expected value $\frac{4}{3}\pi^2$ in the regions of large value of
$N_5$ and ${\cal P}_5$. This indicates that our hypothesis does not require
some preferred value of $M$ and, therefore, some of the configurations with
large $N_5$ are also compatible with the assumption of random distribution
of the orbital phases.}

\section{Conclusions}
\label{sec:conclusions}
The massive molecular torus (CND) surrounding the central parsec of the
Galactic Centre causes precession of the orbits of young stars which move at
distances $0.03\mathrm{pc}\lesssim r \lesssim 0.3\mathrm{pc}$ around the
supermassive black hole. The rate of the precession depends on the orbital
parameters as well as on the orientation and mass of the CND.
This rate is comparable to the lifetime of the young stars for a wide range of
parameters and, therefore, this process should be taken into
consideration in attempts to determine the relation between initial and current
values of their orbital parameters. We have shown that $\mcnd\gtrsim0.3\mbh$
would destroy any coherently rotating disc-like stellar structure within
$6\mathrm{Myr}$, provided the inclination of most of the orbits with respect
to the CND deviates by more than $5\degr$ from $90\degr$. In other words, the
stability of the stellar disc within its lifetime poses constraints on its
inclination with respect to the CND and on the mass of the CND.

We further suggest that {\em most if not all\/} young stars observed in the
Galactic Centre may have been formed in a single, initially coherently rotating
structure, presumably via fragmentation of a thin self-gravitating gaseous disc.
Within this hypothesis, the orientation of the stellar disc was nearly exactly
perpendicular with respect to the CND. Its `core', represented by the CWS
nowadays, remained nearly untouched by the precession. On the other hand, stars
that were formed at the outer parts of the disc and/or slightly off its mean
plane, or that were scattered out of it via two-body encounters,
have undergone a more rapid precession of their
orbits, i.e. they apparently don't belong to the stellar disc any longer.
We have shown that within the $1\sigma$ uncertainty of their current velocities
there exist such parameters of the stellar orbits that would have had their
angular momenta collinear about $6\mathrm{Myr}$ ago. Due to the high
sensitivity of the precession upon the orbit inclination with respect to the
CND and the uncertainty in the observed parameters of the stellar orbits, the
procedure described in the previous Section cannot provide robust constraints
on the parameters of the model. Therefore, the concept of a single warped disc
of young stars in the Galactic Centre may be considered as being viable, but
not proven yet. Our hypothesis, however, gives an explicit prediction on
a specific pattern of the normal vectors of the stellar orbits which may be
determined from future, more accurate observations: all of them are assumed
to be found close to the circumference perpendicular to the normal vector
of the CND.

Let us emphasise that the gravitational influence of the CND leaves stronger
imprints on the dynamics of stars more distant from the centre. Hence, we
suggest that these stars deserve further attention from the observational
point of view. Improved measurements of their kinematical state may bring
a new light into the question of the formation and dynamical evolution of
the population of the young stars in the Galactic Centre.
Beside a generic demand on better constraints on orbital parameters from the
observational side, there is also a room for improvements of
the model itself. Its most important (and computationally rather expensive)
modification will probably lie in an improved treatment of the evolution of
the individual orbits, which would take into account gravity of the stellar
disc itself.

As a final remark let us note that the strict constraints on the mutual
(perpendicular) orientation of the stellar disc and the CND raises a question
about the dynamics of gas from which the young stars were formed. It is likely
that the parent gaseous disc had to be nearly perpendicular to the CND, so
that it would not be destroyed via differential precession before it gave birth
to the numerous stellar population. Such an initial orientation is statistically
not very probable, opening the question whether it can be a generic result of
dissipative (hydro)dynamics in the resonant external potential.

\section*{Acknowledgements}
We thank an anonymous referee for helpful comments.
L.\v{S}. gratefully appreciates a fellowship from the Alexander von Humboldt
Foundation and the hospitality of the host institute (AIfA). This work was
also supported by the DFG Priority Program 1177,
the Research Program MSM0021620860 of the Czech Ministry of Education
and the Czech Science Foundation (ref.\ 205/07/0052).
J.S. is grateful to acknowledge utilisation the Grid infrastructure 
of the Enabling Grids for E-sciencE (EGEE II) project, a project co-funded by the
European Commission under contract number INFSO-RI-031688.


\end{document}